\documentclass[twocolumn,showpacs,preprintnumbers,amsmath,amssymb]{revtex4}

\usepackage{graphicx}% Include figure files
\usepackage{dcolumn}% Align table columns on decimal point
\usepackage{bm}% bold math

\begin{document}

\title{Scaling and memory in recurrence intervals of Internet traffic}
\author{Shi-Min Cai$^{1,2}$}
\author{Zhong-Qian Fu$^{1}$}
\author{Tao Zhou$^{2,3}$}
\email{zhutou@ustc.edu}
\author{Jun Gu$^{1}$}
\author{Pei-Ling Zhou$^{1}$}

\affiliation{$^{1}$ Department of Electronic Science and Technology,
University of Science and Technology of China, Hefei Anhui, 230026,
PR China\\ $^{2}$Department of Physics, University of Fribourg,
Chemin du Mus\'ee 3, 1700 Fribourg, Switzerland
\\$^{3}$Department of Modern Physics, University of
Science and Technology of China, Hefei Anhui, 230026, PR China }

\date{\today}

\begin{abstract}
By studying the statistics of recurrence intervals, $\tau$, between
volatilities of Internet traffic rate changes exceeding a certain
threshold $q$, we find that the probability distribution functions,
$P_{q}(\tau)$, for both byte and packet flows, show scaling property
as
$P_{q}(\tau)=\frac{1}{\overline{\tau}}f(\frac{\tau}{\overline{\tau}})$.
The scaling functions for both byte and packet flows obeys the same
stretching exponential form, $f(x)=A\texttt{exp}(-Bx^{\beta})$, with
$\beta \approx 0.45$. In addition, we detect a strong memory effect
that a short (or long) recurrence interval tends to be followed by
another short (or long) one. The detrended fluctuation analysis
further demonstrates the presence of long-term correlation in
recurrence intervals.
\end{abstract}


\maketitle

Many complex systems are characterized by heavy-tailed
distributions, such as power-law distributions \cite{Newman2005},
lognormal distributions \cite{Crow1988}, and stretched exponential
distributions \cite{Laherrere1998}. These distributions imply a
nontrivial probability of the occurrences of the extreme events.
Statistical laws on these extreme events provides evidence for the
understanding of the mechanism that underlies the dynamical
behaviors of the corresponding complex systems. Recently, some
typical complex systems, such as earthquakes
\cite{Bak,Corral1,Corral2,Corral3,Livina}, financial markets
\cite{Wang2001,Lillo,Yamasaki,Wang} and many other natural hazards
\cite{Bunde}, have been widely investigated. Taking the earthquakes
for example, excluding the well established \emph{Omori Law}
\cite{Omori} and \emph{Gutenberg-Richter Law} \cite{Gutenberg}, the
scaling law for temporal and spatial variability of earthquakes have
been observed by Bak \emph{et al}. \cite{Bak} and Corral
\cite{Corral1,Corral2,Corral3}, and the memory effect in the
occurrence of the earthquakes is revealed by showing the statistics
of the recurrence times above a certain magnitude \cite{Livina}.

The Internet has been viewed as a typical complex system that
evolves in time through the addition and removal of nodes and links,
and empirical evidence has demonstrated its small-world and
scale-free structural properties \cite{Faloutsos3,NJP}. One of the
research focuses, the Internet traffic, has been widely studied by
computer scientists, physicists and beyond. For instance, Leland
\emph{et al.} first found the self-similar nature and long-range
dependence of Ethernet traffic that have serious implications for
the design, congestion control, and analysis of computer
communication networks \cite{Leland}. After that, several traffic
models are proposed to understand the underlying mechanism for
information transport and congestion control of the Internet traffic
\cite{Taqqu6,Taqqu9,Taqqu10}, especially, those models (see Refs.
\cite{Taqqu6,Taqqu9} about the models and the Ref. \cite{Taqqu10}
about the time series analysis) can, to some extent, reproduce the
self-similar nature of the Internet traffic, which indicates the
existence of burstiness of traffic and the large volatility of
traffic rate changes.

\begin{figure}
\scalebox{0.8}[0.8]{\includegraphics{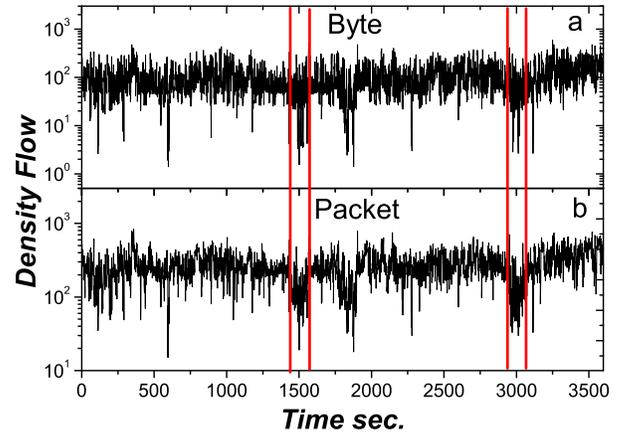}}
\caption{\label{fig:epsart}(Color online) The density flows of both
bytes and packets in second resolution. Between red lines, the
clustering behaviors are observed.}
\end{figure}

\begin{figure}
\scalebox{0.8}[0.8]{\includegraphics{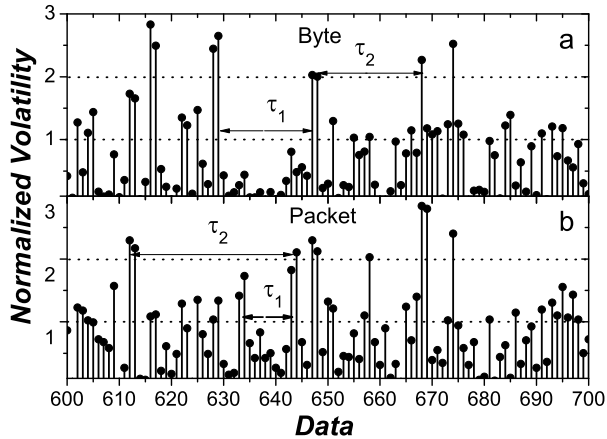}}
\caption{\label{fig:epsart} \textbf{Illustration of recurrence
intervals, $\tau_q$, of normalized volatility time series with $q=1$
and $q=2$.}}
\end{figure}

\begin{figure}
\scalebox{0.7}[0.8]{\includegraphics{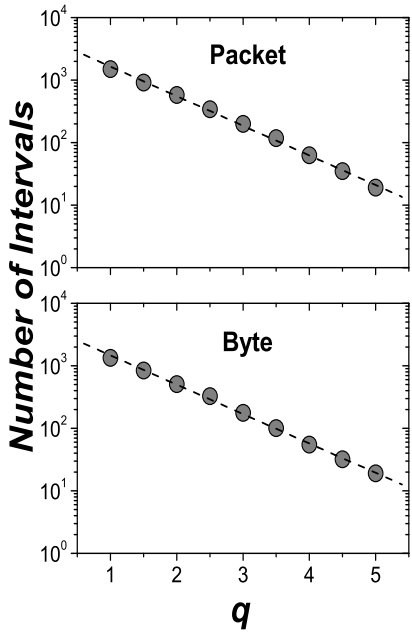}}
\caption{\label{fig:epsart} \textbf{Number of intervals vs. the
threshold $q$ in linear-log plots. Both of the two fitting lines are
of slopes 0.474.}}
\end{figure}

\begin{figure} \scalebox{0.7}[0.8]{\includegraphics{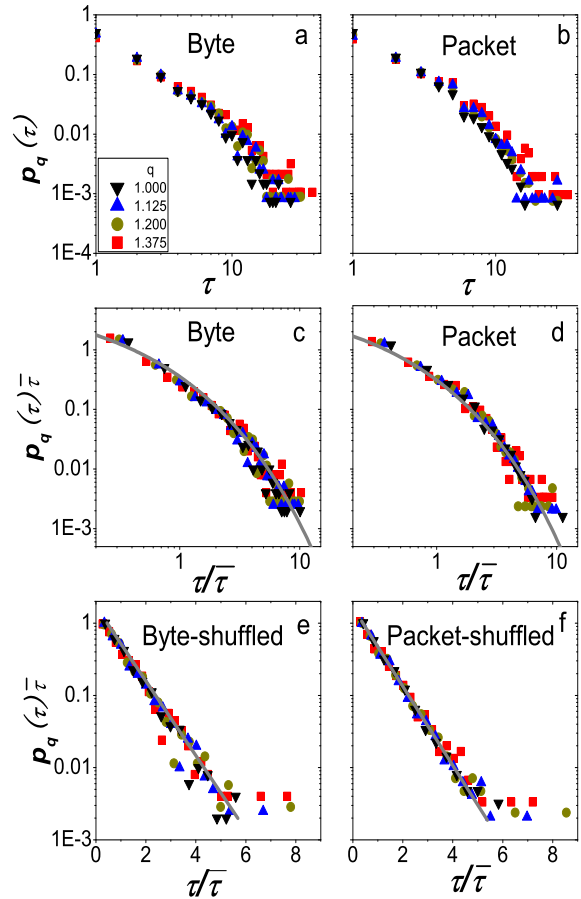}}
\caption{\label{fig:epsart}(Color online) (a) and (b): Distributions
of recurrence intervals between two consecutive volatilities of
traffic rate changes above some thresholds. (c) and (d): Rescaled
distributions $P_{q}(\tau)\overline{\tau}$ vs. rescaled recurrence
intervals $\frac{\tau}{\overline{\tau}}$. The solid lines denote a
stretching exponential function $f(x)=A\texttt{exp}(-Bx^{\beta})$
with $\beta \approx 0.45$. (e) and (f): Rescaled distributions
$P_{q}(\tau)\overline{\tau}$ vs. $\frac{\tau}{\overline{\tau}}$ for
shuffled data, which obey an exponential form
$f(x)=A\texttt{exp}(-Bx)$. \textbf{The rescaled distributions of
real data have heavier tails than the ones of shuffled data.}}
\end{figure}

Herein we are interested in the large volatility that implies the
suddenly drastic changes of traffic rate. In previous studies, by
analyzing a set of time series data of round-trip time, Abe and
Suzuki \cite{Abe1,Abe2} reported that the drastic changes, named of
\emph{Internet quakes}, are characterized with the Omori Law and
Gutenberg-Richter Law. By statistical analysis on recurrence
interval $\tau$ between the volatility of traffic rate changes
exceeding a certain threshold $q$, this Letter reports that:
(\emph{i}) the probability distribution functions (pdfs)
$P_{q}(\tau)$ for both byte and packet flows, rescaled by the mean
recurrence interval $\overline{\tau}$, yield scaling property that
the scaling function $f(x)$ follows a stretching exponential form
$f(x)=A\texttt{exp}(-Bx^{\beta})$ with $\beta \approx 0.45$ for all
data; (\emph{ii}) a short/long recurrence interval tends to be
followed by another short/long recurrence interval, implying a
strong memory effect.

The data used in this paper are part of Ethernet traffic set
collected at Bellcore. They correspond to one ``normal" hour's worth
of traffic, collected every 10 milliseconds, hence resulting in a
time series with a length of 360000. There are two types of
measurements, recording the number of bytes and packets per unit
time, respectively. The data and information can be found at the
Internet traffic archive \cite{web2}. These data have been widely
used and become the most important benchmark data in relevant areas.
We firstly integrate the time series into a second resolution, that
is, each integrated data point is an average of 100 original points.
As shown in Fig. 1, the time series exhibit clustering phenomenon
that is resulted from the traffic congestion in the Internet. For
both byte and packet flows, we use the absolute value of changes,
$|\Delta S_{i}|=|S_{i}-S_{i-1}|$ where $S_i$ denotes the data point
at time $i$, to quantify the volatility. It has been demonstrated
that the pdf of $\Delta S_{i}$ decays in an asymptotic power-law
form and the volatilities are long-term correlated \cite{Cai}. We
then normalize the volatility time series by the standard deviation
$(\langle |\Delta S_{i}|^{2} \rangle - \langle| \Delta S_{i}|
\rangle ^{2})^{1/2}$. In this way, the threshold $q$ are in units of
the standard deviation of the volatility. An illustration of
recurrence interval $\tau$ is shown in Fig. 2.

\begin{figure}
\scalebox{0.7}[0.8]{\includegraphics{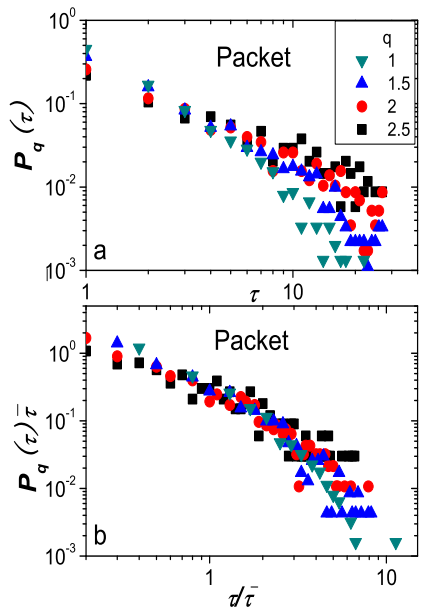}}
\caption{\label{fig:epsart}(Color online) \textbf{Original
distribution (a) and the rescaled distribution (b) of recurrence
intervals between two consecutive volatilities for packet flow.
Here, the range of threshold $q$ is much larger than that in the
Figure 4. The case for byte flow is almost the same, thus it is
omitted here.}}
\end{figure}

\begin{figure}
\scalebox{0.8}[0.8]{\includegraphics{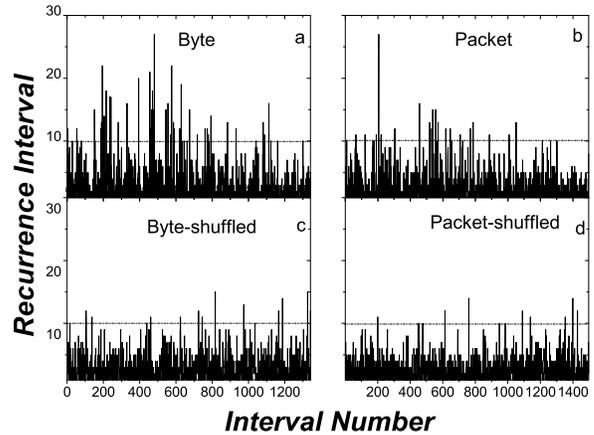}}
\caption{\label{fig:epsart}(a) and (b): Typical examples of
recurrence intervals for byte and packet flows with $q=1$. (c) and
(d): Same as (a) and (b), except that the original volatility time
series are shuffled. The horizontal line is used to indicate the
cluster of large recurrence intervals.}
\end{figure}

\begin{figure}
\scalebox{0.8}[0.8]{\includegraphics{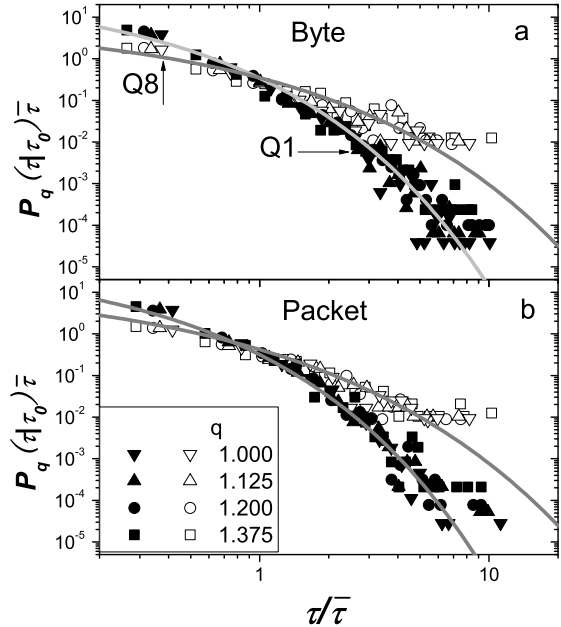}}
\caption{\label{fig:epsart} Conditional pdfs,
$P_{q}(\tau|\tau_{0})$, of recurrence intervals of byte and packet
flows for $\tau_{0}$ in $Q_{1}$ (full symbols) and $Q_{8}$ (open
symbols) as functions of $\frac{\tau}{\overline{\tau}}$. The fitting
curves follow the stretched exponential forms.}
\end{figure}

\textbf{For the normalized data, the smaller value of threshold,
$q$, does not suggest the recurrence interval between the large
volatilities (extreme events), therefore we concentrate on the cases
with $q\geq 1$. As shown in Fig. 3, for both byte and packet flows,
the number of recurrence intervals decays exponentially fast as the
increasing of $q$. For large $q$, the results are inaccurate and
incredible for the low statistics, and thus we mainly discuss the
statistics for a very limited range of $q$.} Figure 4(a) and 4(b)
present the behaviors of the pdfs, $P_{q}(\tau)$, for both byte and
packet flows with different $q$, which are obviously broader than
the Poisson distributions as for uncorrelated data, and the pdf for
larger $q$ decays slower than that for smaller $q$. To understand
how $P_{q}(\tau)$ depends on $q$, in Fig. 4(c) and 4(d), we show the
rescaled pdfs, $P_{q}(\tau)\overline{\tau}$, for byte and packet
flows as functions of the rescaled recurrence intervals
$\frac{\tau}{\overline{\tau}}$. The data collapse to a single curve,
indicating a scaling relation
\begin{equation}
P_{q}(\tau)=\frac{1}{\overline{\tau}}f(\frac{\tau}{\overline{\tau}}),
\end{equation}
which suggests that the scaling function does not directly depend on
the threshold $q$ but only through
$\overline{\tau}=\overline{\tau}(q)$. Furthermore, as shown in Fig.
4(c) and 4(d), the scaling functions for both byte and packet flows
follow the same stretching exponential form,
$f(x)=A\texttt{exp}(-Bx^{\beta})$ with $\beta \approx 0.45$,
indicating a possibly universal scaling property in the recurrence
intervals of Internet traffic data \textbf{(see also a similar
scaling in the intertrade time in financial markets
\cite{Ivanov2004})}. Therefore, one can estimate $P_{q}(\tau)$ for
an arbitrary $q$ with the knowledge of $P_{q'}(\tau)$ for a certain
$q'$. This scaling property is particularly significant for the
understanding of the statistics of large-$q$ case where the number
of data is usually very small. \textbf{Similar analysis for very
large $q$ has also been done, as shown in Fig. 5, the rescaled
distributions get much closer to each other than the original
distributions. However, it is hard to tell whether these curves
collapse a single master curve since the number of intervals for
large $q$ is very small. Hereinafter, we focus on the statistics for
$q \in [0,1.375]$.}

The stretching exponential distributions of rescaled recurrence
intervals suggest the existence of correlation of volatilities. In
contrast, the recurrence intervals for uncorrelated time series are
expected to follow a Poisson distribution, as
$\texttt{log}f(x)\sim-x$. To confirm this expectation, the
volatilities are shuffled to remove the correlations, and the
resulting distributions, as shown in Fig. 4(e) and 4(f), decay in an
exponential form, which is remarkably different from that of the
real time series. Furthermore, the very short and very long
recurrence intervals occur more frequently in the real data (see
Fig. 4(c)-4(f)), indicating a burstiness of Internet traffic,
similar as observed in many other complex systems \cite{Goh2008}.

\begin{figure}
\scalebox{0.8}[0.8]{\includegraphics{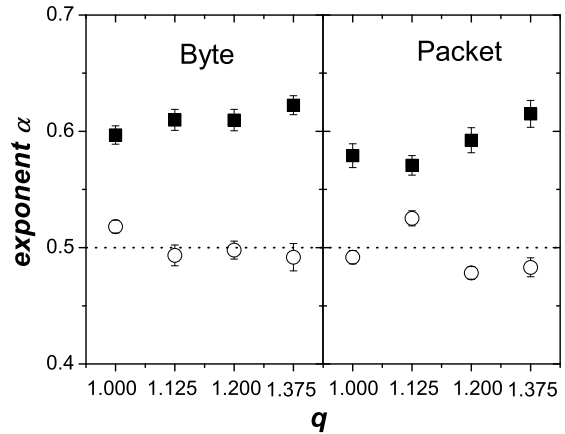}}
\caption{\label{fig:epsart} Long-term memory in recurrence intervals
of byte and packet flows. The full and open symbols represent the
exponent $\alpha$ for the real data and the shuffled data,
respectively. \textbf{The mean value of $\alpha$ for the real data
is $0.600$ with mean error $0.009$, while for the shuffled data it
is $0.497$ with mean error $0.007$.}}
\end{figure}

The scaling property of $P_{q}(\tau)$ of recurrence intervals only
indicates the long-term correlations of volatility time series of
traffic rate changes, but does not tell if the recurrence intervals
are themselves correlated. To answer this question, we next
investigate the memory effect in the recurrence intervals. Before
that, we show a typical example of recurrence intervals for both
byte and packet flows in Fig. 6(a) and 6(b), as well as the
corresponding shuffled sequences in Fig. 6(c) and 6(d). Compared
with the shuffled ones, the original data exhibit the clustering of
large intervals, which indeed indicates the memory effect that a
short (or long) recurrence interval tends to be followed by another
short (or long) one.

To quantify the memory effect, we study the conditional pdf
$P_{q}(\tau|\tau_{0})$, representing the probability a recurrence
interval, $\tau$, immediately follows a recurrence interval,
$\tau_{0}$. If no memory effect exists, $P_{q}(\tau|\tau_{0})$ will
be identical to $P_{q}(\tau)$ and independent to $\tau_{0}$.
Therefore, we study $P_{q}(\tau|\tau_{0})$ not for a specific
$\tau_{0}$, but for a range of $\tau_{0}$ values. Analogous to the
analysis of daily volatility return intervals \cite{Yamasaki}, the
data set of recurrence intervals are sorted in increasing order and
divided into eight subsets, $Q_{1}$, $Q_{2}$, $\cdots$, $Q_{8}$, so
that each subset contains $1/8$ of the total data. It makes the
$N/8$ lowest recurrence intervals belong to $Q_{1}$, whereas the
$N/8$ largest ones belong to $Q_{8}$, where $N$ denotes the total
number of data points. Figure 7(a) and 7(b) show
$P_{q}(\tau|\tau_{0})$ for byte and packet flows. The distribution
corresponding to $Q_1$ if obtained by recording all the $\tau$
values (they form a distribution) if their predecessor, $\tau_0$ is
no less then the smallest interval in $Q_1$ and no more than the
largest interval in $Q_1$ (see Ref. \cite{Yamasaki} for more
details). As shown in Fig. 7, the rescaled pdfs,
$P_{q}(\tau|\tau_{0})\overline{\tau}$ for different $q$, collapse
into a single curve that can also be fitted by stretching
exponential functions. The remarkable difference between the
distributions with $\tau_0$ in $Q_1$ and $Q_8$ clearly demonstrated
the existence of memory effect.

To check whether the memory effect is limited only to the
neighboring recurrence intervals, we use the detrended fluctuation
analysis (DFA), which is a benchmark method to quantify long-term
correlations \textbf{(see Ref. \cite{Peng1994} for the original
method, as well as Ref. \cite{Hu2001} and Ref. \cite{Chen2002} for
the effects of trends and nonstationarities, respectively)}. The
fluctuation $F(l)$ of a time series and window of $l$ seconds,
computed by DFA, follows a power-law relation as $F(l)\sim
l^{\alpha}$. Uncorrelated time series corresponds to $\alpha=0.5$,
while the larger (or smaller) $\alpha$ indicates long-term
correlation (or anti-correlation). \textbf{Figure 8 shows the values
of $\alpha$ for recurrence intervals, which are all larger than 0.5
and of which mean value is $0.600$ with mean error $0.009$,
indicating the presence of long-term correlations in recurrence
intervals. Furthermore, for the shuffled recurrence intervals, the
long-term correlations are absent with $\alpha \simeq 0.5$ (mean
value is $0.497$ with mean error 0.007).}

In summary, we have investigated the scaling and memory properties
in recurrence intervals of the Internet traffic. The empirical pdfs
$P_{q}(\tau)$ for byte and packet flows, respectively, can fall into
a single curve by rescaling with the mean recurrence intervals
$\overline{\tau}$, as shown in Eq. (1). The scaling function has a
stretching exponential form, as $f(x)=A\texttt{exp}(-Bx^{\beta})$
with $\beta \approx 0.45$ for both byte and packet flows. This
scaling property can be used to predict the occurrence probability
of on rare events that correspond to large $q$. We also detected the
memory effect that a short (or long) recurrence interval tends to be
followed by another short (or long) one, which is further
demonstrated by the empirical results that the conditional pdf,
$P_{q}(\tau|\tau_{0})$, is strongly dependent on $\tau_0$. Further
more, by using the DFA method, we found that the recurrence
intervals are indeed long-term correlated. Some recently reported
empirical studies show that the Internet-based human activities
exhibit burstiness and memory in temporal statistics, such as the
web accessing \cite{Dezso2006,Goncalves2008} and on-line
entertainment \cite{Zhou2008}. All those activities contribute some
to the Internet traffic, and thus we think the analysis of the
burstiness and memory of the Internet traffic itself can be
considered as a valuable complementary work. More interestingly, the
results suggest that the Internet shares some common properties with
other complex systems like earthquake and financial market
\cite{Wang2001,Lillo,Yamasaki,Wang}, which gives support to the
possibly generic organizing principles governing the dynamics of
apparently disparate complex systems, as dreamed by Goh and
Barab\'asi \cite{Goh2008}.

This work is supported by the National Science Foundation of China
under Grant Nos. 70671097, 60874090 and 10635040. S.-M.C.
acknowledges the financial support of State 863 Projects
(2008AA01A318).

\end{document}